The design of a very high-Q superconductor electromechanical clock.

Osvaldo F Schilling

Departamento de Física, Universidade Federal de Santa Catarina, Campus, Trindade, 88040-900, Florianópolis, SC. Brazil.

PACS numbers: 85.25.-j , 84.30.Ng , 84.60.-h

**Abstract.** We discuss theoretically the properties of an electromechanical oscillator whose operation is based upon the cyclic, quasi-conservative conversion between gravitational potential, kinetic, and magnetic energies. The system consists of a strong-pinning type-II superconductor square loop subjected to a constant external force and to magnetic fields. The loop oscillates, and has induced in it a rectified electrical current. The emphasis of this paper is on the evaluation of the major remaining source of losses in the oscillations. We argue that such losses should be associated with the viscous vibration of pinned flux lines in the superconductor Nb-Ti wire, provided the oscillator is kept under vacuum and the magnetic field is sufficiently uniform. We discuss how other different sources of loss would become negligible for such operational conditions, so that a very high quality factor Q exceeding $10^{10}$ might in principle be reached by the oscillator. The prospective utilization of such oscillator as a low-frequency high-Q clock is analyzed. Since publication the ideas in this paper have been explored both by the author and elsewhere, in applications covering Metrology, quantum systems, and gravimetry.

1. Introduction.

In a previous paper [1] we discussed the physical principles of a superconducting electromechanical oscillator whose most remarkable property would be the possibility of isolating, and practically eliminating almost every source of energy loss. Its operation principle is based upon the cyclic and quasi-conservative energy conversion between the magnetic energy associated with the lossless electric currents inside a superconducting loop and the kinetic and gravitational potential energies of the vertical oscillations of the loop in the presence of a system of magnets. The magnetic levitation of the loop itself is not obtained by the usual effect of hysteretic repulsion against a dipolar magnetic field, like in other applications[2], but comes straight from uniform magnetic fields through the Lorentz forces they impose upon the currents in the loop. Electrical currents in the loop are obtained from mechanical motion and vice-versa. It is the objective of the present paper to show theoretically the extreme stability of such oscillator by performing a calculation of its quality factor Q, something not included in [1]. We further discuss possible applications of the oscillator in Metrology and other subjects developed in recent years.

The layout of the oscillator is depicted in Figure 1[1]. Under certain experimental conditions a type-II superconductor square loop of side length $l$, mass $m$, and self-inductance $L$ would levitate in the upright position while subjected to the uniform distribution of fields $B_1$ and $B_2$. The length of the lower side of the loop subjected to $B_2$ is designated as $a$ ($< l$). According to the theory in [1], in the "absence" of losses the loop would perform quasi-harmonic oscillations of frequency $\Omega = B_0 a/(mL)^{1/2}$, with $B_0 \equiv B_1 - B_2$. The amplitude of the oscillations is $x_0 = g/\Omega^2$, and the maximum speed of the oscillations is $v_0 = g/\Omega$, where $g$ stands for the gravity acceleration (assuming $F = mg$ in [1]). Such mechanical oscillations are accompanied by a rectified current containing an alternating component of same frequency, and amplitude $i_0 = mg/(aB_0)$. In practice, the loop should be fixed to a nonmagnetic (e.g., teflon) frame to increase rigidity and avoid vibrations of the thin wire, so that $m$ would stand for the mass of the whole oscillating set.

2. Theory.

Working with a type-II superconductor loop above the lower critical field $H_{c1}$ avoids the problems arising from the normal-state resistivity of the intermediate state regions that might arise if a type-I superconductor material is adopted. However, this introduces several other sources of loss, which must be individually circumvented. For this purpose the loop should oscillate in a vacuum and electrically insulating magnets should be used, to avoid friction against an atmosphere, and to avoid eddy current losses induced in nearby metallic parts. Losses in the superconductor itself are practically eliminated by working well below $T_c$ and by keeping $\Omega$ well below the MHz range, since no eddy currents associated with the normal electrons in the superconductor would be excited at these low frequencies[3]. A strong-pinning, high-$\kappa$ ($\kappa$ is the Ginzburg-Landau parameter) material should be used to make the wire, to avoid flux-creep and hysteresis losses. Flux-creep occurs when bundles of pinned flux-lines (FL) jump between pinning-centers inside a type-II superconductor, under the effect of magnetic forces associated with the current density that flows in the material. For conventional (i.e., low $T_c$) superconductors flux-creep becomes negligible if the current density is much smaller than the critical current density[4], so that the experiment must be designed to obey this condition ( see below).

Hysteresis is by far the most important source of energy dissipation in strong-pinning superconductors carrying alternating currents at low frequencies, under external magnetic fields. Hysteretic losses are associated with the magnetic work per cycle needed to rebuild the FL array, whose displacements are essentially irreversible. Therefore, the key for eliminating hysteresis losses is not allowing the FL in the wire to move irreversibly between pinning centers. Such displacements of the FL are associated with the establishment of a critical-state flux profile in the specimen. We recall some results of Campbell's work on this subject[5,6] quoted in [1]. Campbell demonstrated that if a type-II superconductor sample is subjected to a static external field $H$ superimposed to a small ripple field $h$ of frequency $\omega$, a critical-state flux profile will be established from the surface only if $b = \mu_0 h$ is greater than $\delta \equiv (\mu_0 J_c B d_o)^{1/2}$, where $J_c$ is the critical current density of the material, $B \approx \mu_0 H$, and $d_o$ is the size of the potential energy well that arrests a flux line ($d_o$ is measured in [6] and is on the order of 2 to 6 nm). For lower ripple fields the FL displacements are quasi-reversible ( i.e., no losses were *measured*). In such conditions the ripple field and the corresponding induced current will be restricted to a surface layer of thickness $\lambda_{\text{eff}} = (B d_o/(\mu_0 J_c))^{1/2}$, which is on the order of micrometers for strong-pinning materials and $B$ on the order of Tesla. $\lambda_{\text{eff}}$ is analogous to the penetration depth of type-I superconductors, but much greater in magnitude. Therefore, to produce reversible FL displacements only the current $i$ flowing in a square loop made with a cylindrical superconductor wire of radius $r$ should not generate a field greater than $\delta$ at any point on the surface of the wire, and thus $i$ must be restricted by the criterion [1]

$$\gamma \mu_0 i/(2\pi r) \leq (\mu_0 J_c B d_o)^{1/2} \tag{1}$$

Here $\gamma \approx 1.75$ corrects for the concentration of field at the inner corners of the loop as compared with the field $\mu_0 i/(2\pi r)$ produced by an infinite wire of same radius. It must be stressed that under such conditions the critical state is not reached since the current density is too small, and this avoids flux-creep also, in agreement with Campbell's observation that no effect of flux-creep was apparent in his data [5,6]. Another effect discarded by Campbell was the hysteretic motion of FL pinned by surface roughness in the specimen, a subject qualitatively discussed by Melville[7]. Campbell's results were fully explainable utilizing bulk-material properties only, with no intervention of surface effects. This indicates that the losses associated with surface pinning are comparatively small. In the discussions below we follow Campbell and consider that surface effects may be eliminated by using high-$\kappa$, strong-pinning materials [6] and by doing a proper surface finishing on the wire.

On the other hand, Campbell's theoretical model completely neglects energy dissipation associated with the FL oscillations inside the pinning wells. This is an oversimplification, since the motion of the FL is accompanied by dragging of the normal electrons in their centers, so that a viscous force term proportional to the velocity of the FL should be included in the theory [8]. In fact, Campbell´s description of the quasi-reversible penetration of flux into a planar vacuum/superconductor interface can be deduced from the ordinary treatment of eddy currents in metals[9] provided one assumes an imaginary resistivity $\rho_C = (i\omega\eta/k)(H/H_{c2})\rho_n$, which takes the viscosity effects partially into account. Here the notation of [8] ( translated to MKS units) is adopted: $\eta = \phi_0\mu_0 H_{c2}/\rho_n$ is the flux-flow viscosity, $k = J_c\phi_0/d_0$ is the elastic force constant of the pinning interaction, $\rho_n$ is the normal-state resistivity, $H_{c2}$ is the upper critical field, and $\phi_0$ is the flux quantum. With such expression for $\rho_C$ Campbell's effective penetration depth $\lambda_{eff}$ may be obtained at once from the formula for the eddy currents skin depth[9] as $\lambda_{eff} = (\mu_0\omega/|\rho_C|)^{-1/2}$. For this purely imaginary resistivity the phasors **J** and **E**$= \rho_C$**J** become 90° out of phase and thus no power $P = \text{Re} \langle \mathbf{E^*J} \rangle$ related to flux motion is dissipated within this level of approximation.

The more precise treatment of Gittleman and Rosenblum[8] results in a resistivity $\rho_{GR} = (i\omega\eta/(i\omega\eta + k))(H/H_{c2})\rho_n$. There will be losses associated with the real part of $\rho_{GR}$. Since the ratio $\eta/k$ is usually much smaller than $10^{-6}$ these losses become extremely small, but not null at low frequencies. Such treatment of viscous-flow losses associated with the motion of FL at low currents can be directly applied to the evaluation of the quality factor $Q$ for the type-II superconductor electromechanical oscillator, since it provides a method for calculating the power losses due to vibrations of pinned FL. We take as a realistic example the losses that would be associated with the oscillations of a loop made with a Nb-48 Wt % Ti alloy described in detail in [10,11] The physical properties relevant for the calculations were all carefully determined for this material, so that a quite precise application of the theory is possible.

3. Quality factor calculation for a strong-pinning Nb-Ti loop.

Meingast and collaborators[10,11] performed a very detailed study of the physical and microstructural properties of a Nb-48 Wt. % Ti alloy containing a homogeneous distribution of $\alpha$-Ti precipitates. In the present calculations we take their data for the particular case of a wire drawn to a 0.645 mm diameter. The upper critical field $\mu_0 H_{c2}$= 11.5 T at 4.2 K; the normal state resistivity at the critical temperature (9.5 K) is $\rho_n$= 7x10$^{-7}$ ohm.m; the critical current density $J_c$ = 3x10$^9$ A/m$^2$ at 4.2 K, under a magnetic field of 0.6 Tesla. It will be assumed that the loop is a square of side length $l$= 5 cm, and $a$=3 cm in Figure 1. Since the average density of this Nb-Ti alloy is 6.6 g/cm$^3$, the loop mass would be $m$= 0.431 g. The self-inductance of the loop $L$ can be calculated from a specific formula, so that $L$= 1.5x10$^{-7}$ H. The static fields produced by the magnets will be assumed as $B_1$= 0.6 T and $B_2$= 0.3 T, so that $B_0$= 0.3 T. In order that hysteresis losses be avoided one of the conditions to be met is that the static fields acting upon each point of the wires must remain constant to a precision of about 10% of the ripple field produced by the currents in the loop, irrespective of the loop vertical displacements. From the value for δ ( see below) this tolerance range should be on the order of $4 \times 10^{-4}$ T. All the parameters needed by the theory can be calculated from these data. We obtain $\Omega$= 1119 rad/s, corresponding to a low frequency of oscillation, $f$ = 178 Hz. The amplitude of oscillation is $x_0$= 7.8 µm, which is quite small. From relation (1) it is possible to calculate the threshold ripple field δ, which is $3.4 \times 10^{-3}$ T, for $B = B_1 =$ 0.6 T and $d_0$= 5 nm in (1) ($d_0$ should be similar to the coherence length measured for the alloy at 4.2 K, $\xi$= 5 nm), which results in the threshold current $i_{th}$= 3.1 A. That is, the loop will oscillate without hysteresis losses provided the current induced by the movement is smaller than $i_{th}$. This is actually the case, since the maximum value of the rectified current necessary for levitation is[1] $i_{max}$= $2i_0$= $2mg/(aB_0)$= 0.94 A, which guarantees that the critical state conditions are far from being reached.

We conclude that if the loop oscillates in a vacuum, the only "major" source of energy dissipation will be the viscous drag of FL oscillating inside their pinning wells in a surface layer of thickness $\lambda_{eff}$. According to [8], the power dissipated per unit volume is $P = \frac{1}{2}(\omega\eta)^2/((\omega\eta)^2 + k^2) J^2 (H/H_{c2})\rho_n$, where we neglect the FL mass and $\omega = \Omega$. Inserting the figures into the expressions for the viscosity and for the elastic constant one obtains $\eta = 3.29 \times 10^{-8}$ and $k = 1200$ (all in MKS units), which makes the $\eta/k$ ratio extremely small. This leads to a simplification in the expression for $P$, which can be written as

$$P_{FL} = \tfrac{1}{2} (\Omega \eta / k)^2 J^2 (H/H_{c2}) \rho_n \qquad (2)$$

The currents in the loop flow within a surface layer of thickness $\lambda_{eff} = 0.63$ μm for $B_2 = 0.3$ T, and 0.89 μm, for $B_1 = 0.6$ T (the same $J_c$ is used for both fields), so that the effective cross-sectional area of wire penetrated by currents is $S_s = 2\pi r \lambda_{eff} = 1.27 \times 10^{-9}$ m$^2$ if the field is $B_2$, and $1.79 \times 10^{-9}$ m$^2$ if the field is $B_1$. The current density in (2) is obtained from the amplitude of the alternating part of the rectified current, that is, $J = i_0/S_s$, for each value of $S_s$. Therefore, the average value of $P_{FL}$ from (2) is $1.2 \times 10^{-6}$ W/m$^3$. Such power is dissipated in a thin tubular shell where the FL oscillations actually take place around the loop. Taking into consideration the variations in $\lambda_{eff}$, the effective power dissipated in the loop is the product of $P$ times the volume of the tubular shells, so that $P_{eff} = 4.1 \times 10^{-16}$ W.

The quality factor $Q$ is defined as the ratio $U\Omega/P_{eff}$, where $U$ is the total energy of the oscillating system. In this case $U = mgx_0 = Li_0^2 = 3.24 \times 10^{-8}$ J, which results in $Q = 8.8 \times 10^{10} \approx 10^{11}$.

3.1 Further corrections.

In reality, the oscilations will not take place under perfect vacuum and at the absolute zero temperature. The loop will be subject to dragging forces due to the remaining atmosphere inside the cryostat even at high-vacuum[11]conditions.

The main factor affecting dragging is temperature, though. Temperature will drastically affect the cryostat atmosphere viscosity $\eta_g$. The dragging force per unit length acting upon a slender cylinder due to a rarefied gaseous atmosphere of specific mass $\rho$ at low oscillating speeds $v$ transverse to its length is described by the formula[12,13]:

$$F_d = 4\pi\eta_g v/(\ln(\eta_g/(v\rho r)) + \tfrac{1}{2} + 2\ln 2 - 0.577215) \qquad (3)$$

The losses associated with (3) can in fact be much greater than those due to the FL oscillations. However, if we take $T = 0.05$ K, $p = 10^{-8}$ Torr, $\eta_g = 1.5 \times 10^{-10}$ kg/(m s)[14], $r = 0.3$ mm, $v$ (maximum) $= \Omega x_0 = 9$ mm/s, and adopt the ideal gas equation of state for $\rho$ for the He gas, we obtain a maximum force $F_d = 1.7 \times 10^{-12}$ N/m. Each vertical side of the loop is subject to about one half the drag of an horizontal side[15]. For sides ($l$) 5 cm long, the effective power associated with $F_d$ will be given by $P_d = 3l \times (\tfrac{1}{2} F_d v) = 1.15 \times 10^{-15}$ W. This number is about three times the value for $P_{FL}$ due to the FL oscillations calculated earlier. Taking all these losses into account, the conclusion is that the *Q* factor for this oscillator would have an upper bound of about $2\times10^{10}$ under the best "realistic" working conditions.

4. Recent developments and potential applications.

Ever since the publication of this paper in 2007 applications have been proposed for the loop-oscillator. In 2013 the author showed theoretically that precise measurements of the flux trapped insided the loop alongside the measurement of its speed of oscillation may lead to a very accurate measurement of the loops mass[16]. It was pointed out that the method had a potential for application in the Metrology of mass standards in the range of the grams.  The design-scheme was embraced also by G.K.Brennen of Macquarie University, Australia. His group proposed the utilization of three mutually ortogonal rectangular loops in place of a single one. The oscillations of a magnetic sphere over the loops-array would then couple to the ground-state of a quantum dot and tend to help the system to settle in its ground state[17,18]. Applications in precise gravimetry have also been proposed by the same group[19]. In recent work the author has explored the concept of confined current motion in basic physics problems[20].

In resume, superconducting oscillations in the presence of gravity and other external fields appear to have deep theoretical basic-Physics implications, as well as important potential applications in Metrology. In addition, the finer the scale the more important the topological aspects of the problem seem to become. Further investigations must concentrate on such aspects.

Author email: osvaldo.neto@ufsc.br

Figure caption.

Figure 1: A square type-II superconductor loop of side $l$ is subjected to a distribution of uniform static fields $B_1$ and $B_2$ greater than the lower critical field. Note that while the loop oscillates in the vertical direction no part of its wires is subjected to variable static magnetic fields, something essential to avoid hysteresis losses[1].

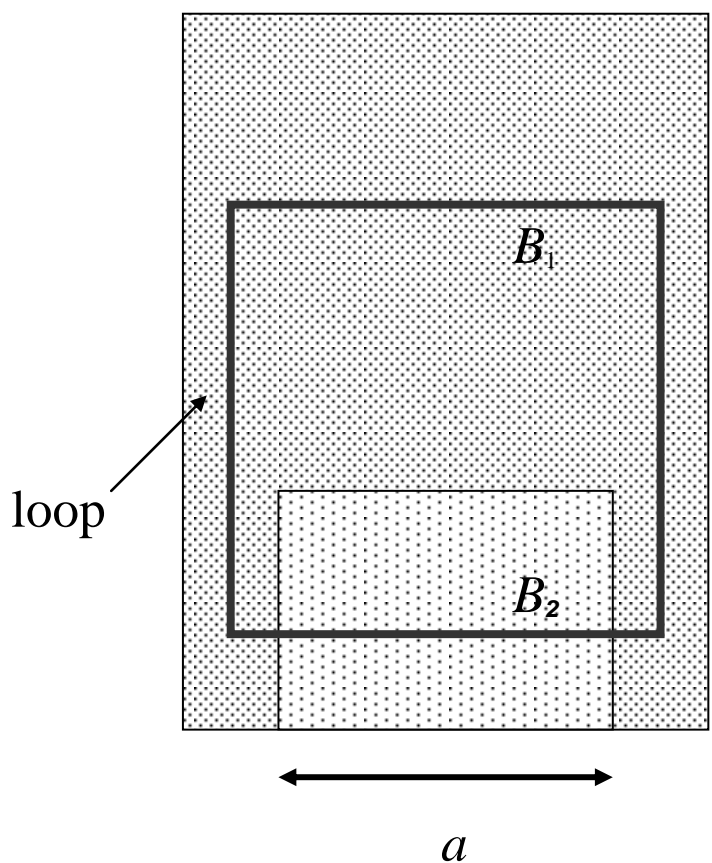

Figure 1